\documentclass[aps,prl,preprint,nopacs,superscriptaddress]{revtex4}
\usepackage{amsmath}
\usepackage{amssymb}
\usepackage{graphicx}
\usepackage{hyperref}
\newcommand{\beq}{\begin{equation*}}
\newcommand{\eeq}{\end{equation*}}
\newcommand{\comp}{Mo$_x$W$_{1-x}$Te$_2$}
\newcommand{\half}{Mo$_{0.45}$W$_{0.55}$Te$_2$}
\newcommand{\ten}{Mo$_{0.07}$W$_{0.93}$Te$_2$}
\newcommand{\twenty}{Mo$_{0.2}$W$_{0.8}$Te$_2$}

\begin{document}

\title{Unoccupied electronic structure and signatures of topological Fermi arcs in the Weyl semimetal candidate \comp}

\author{Ilya Belopolski\footnote{These authors contributed equally to this work.}} \email{ilyab@princeton.edu}
\affiliation{Laboratory for Topological Quantum Matter and Spectroscopy (B7), Department of Physics, Princeton University, Princeton, New Jersey 08544, USA}

\author{Su-Yang Xu$^*$}
\affiliation{Laboratory for Topological Quantum Matter and Spectroscopy (B7), Department of Physics, Princeton University, Princeton, New Jersey 08544, USA}

\author{Yukiaki Ishida$^*$}
\affiliation{The Institute for Solid State Physics (ISSP), University of Tokyo, Kashiwa-no-ha, Kashiwa, Chiba 277-8581, Japan}

\author{Xingchen Pan$^*$}
\affiliation{National Laboratory of Solid State Microstructures, Collaborative Innovation Center of Advanced Microstructures, and Department of Physics, Nanjing University, Nanjing, 210093, P. R. China}

\author{Peng Yu$^*$}
\affiliation{Centre for Programmable Materials, School of Materials Science and Engineering, Nanyang Technological University, 639798, Singapore}

\author{Daniel S. Sanchez}
\affiliation{Laboratory for Topological Quantum Matter and Spectroscopy (B7), Department of Physics, Princeton University, Princeton, New Jersey 08544, USA}

\author{Madhab Neupane}
\affiliation{Department of Physics, University of Central Florida, Orlando, FL 32816, USA}

\author{Nasser Alidoust}
\affiliation{Laboratory for Topological Quantum Matter and Spectroscopy (B7), Department of Physics, Princeton University, Princeton, New Jersey 08544, USA}

\author{Guoqing Chang}
\affiliation{Centre for Advanced 2D Materials and Graphene Research Centre, National University of Singapore, 6 Science Drive 2, 117546, Singapore}
\affiliation{Department of Physics, National University of Singapore, 2 Science Drive 3, 117546, Singapore}

\author{Tay-Rong Chang}
\affiliation{Department of Physics, National Tsing Hua University, Hsinchu 30013, Taiwan}

\author{Yun Wu}
\affiliation{Ames Laboratory, U.S. DOE and Department of Physics and Astronomy, Iowa State University, Ames, Iowa 50011, USA}

\author{Guang Bian}
\affiliation{Laboratory for Topological Quantum Matter and Spectroscopy (B7), Department of Physics, Princeton University, Princeton, New Jersey 08544, USA}

\author{Hao Zheng}
\affiliation{Laboratory for Topological Quantum Matter and Spectroscopy (B7), Department of Physics, Princeton University, Princeton, New Jersey 08544, USA}

\author{Shin-Ming Huang}
\affiliation{Centre for Advanced 2D Materials and Graphene Research Centre, National University of Singapore, 6 Science Drive 2, 117546, Singapore}
\affiliation{Department of Physics, National University of Singapore, 2 Science Drive 3, 117546, Singapore}

\author{Chi-Cheng Lee}
\affiliation{Centre for Advanced 2D Materials and Graphene Research Centre, National University of Singapore, 6 Science Drive 2, 117546, Singapore}
\affiliation{Department of Physics, National University of Singapore, 2 Science Drive 3, 117546, Singapore}

\author{Daixiang Mou}
\affiliation{Ames Laboratory, U.S. DOE and Department of Physics and Astronomy, Iowa State University, Ames, Iowa 50011, USA}

\author{Lunan Huang}
\affiliation{Ames Laboratory, U.S. DOE and Department of Physics and Astronomy, Iowa State University, Ames, Iowa 50011, USA}

\author{You Song}
\affiliation{State Key Laboratory of Coordination Chemistry, School of Chemistry and Chemical Engineering, Collaborative Innovation Center of Advanced Microstructures, Nanjing University, Nanjing, 210093, P. R. China}

\author{Baigeng Wang}
\affiliation{National Laboratory of Solid State Microstructures, Collaborative Innovation Center of Advanced Microstructures, and Department of Physics, Nanjing University, Nanjing, 210093, P. R. China}

\author{Guanghou Wang}
\affiliation{National Laboratory of Solid State Microstructures, Collaborative Innovation Center of Advanced Microstructures, and Department of Physics, Nanjing University, Nanjing, 210093, P. R. China}

\author{Yao-Wen Yeh}
\affiliation{Princeton Institute for Science and Technology of Materials, Princeton University, Princeton, New Jersey, 08544, USA}

\author{Nan Yao}
\affiliation{Princeton Institute for Science and Technology of Materials, Princeton University, Princeton, New Jersey, 08544, USA}

\author{Julien Rault}
\affiliation{Synchrotron SOLEIL, L'Orme des Merisiers, Saint-Aubin-BP 48, 91192 Gif-sur-Yvette, France}

\author{Patrick Lefevre}
\affiliation{Synchrotron SOLEIL, L'Orme des Merisiers, Saint-Aubin-BP 48, 91192 Gif-sur-Yvette, France}

\author{Fran\c{c}ois Bertran}
\affiliation{Synchrotron SOLEIL, L'Orme des Merisiers, Saint-Aubin-BP 48, 91192 Gif-sur-Yvette, France}

\author{Horng-Tay Jeng}
\affiliation{Department of Physics, National Tsing Hua University, Hsinchu 30013, Taiwan}
\affiliation{Institute of Physics, Academia Sinica, Taipei 11529, Taiwan}

\author{Takeshi Kondo}
\affiliation{The Institute for Solid State Physics (ISSP), University of Tokyo, Kashiwa-no-ha, Kashiwa, Chiba 277-8581, Japan}

\author{Adam Kaminski}
\affiliation{Ames Laboratory, U.S. DOE and Department of Physics and Astronomy, Iowa State University, Ames, Iowa 50011, USA}

\author{Hsin Lin}
\affiliation{Centre for Advanced 2D Materials and Graphene Research Centre, National University of Singapore, 6 Science Drive 2, 117546, Singapore} \affiliation{Department of Physics, National University of Singapore, 2 Science Drive 3, 117546, Singapore}

\author{Zheng Liu} \email{z.liu@ntu.edu.sg}
\affiliation{Centre for Programmable Materials, School of Materials Science and Engineering, Nanyang Technological University, 639798, Singapore}
\affiliation{NOVITAS, Nanoelectronics Centre of Excellence, School of Electrical and Electronic Engineering, Nanyang Technological University, 639798, Singapore}
\affiliation{CINTRA CNRS/NTU/THALES, UMI 3288, Research Techno Plaza, 50 Nanyang Drive, Border X Block, Level 6, 637553, Singapore}

\author{Fengqi Song} \email{songfengqi@nju.edu.cn}
\affiliation{National Laboratory of Solid State Microstructures, Collaborative Innovation Center of Advanced Microstructures, and Department of Physics, Nanjing University, Nanjing, 210093, P. R. China}

\author{Shik Shin}
\affiliation{The Institute for Solid State Physics (ISSP), University of Tokyo, Kashiwa-no-ha, Kashiwa, Chiba 277-8581, Japan}

\author{M. Zahid Hasan} \email{mzhasan@princeton.edu}
\affiliation{Laboratory for Topological Quantum Matter and Spectroscopy (B7), Department of Physics, Princeton University, Princeton, New Jersey 08544, USA}
\affiliation{Princeton Institute for Science and Technology of Materials, Princeton University, Princeton, New Jersey, 08544, USA}

\pacs{}

\begin{abstract}
Weyl semimetals have sparked intense research interest, but experimental work has been limited to the TaAs family of compounds. Recently, a number of theoretical works have predicted that compounds in the \comp\ series are Weyl semimetals. Such proposals are particularly exciting because \comp\ has a quasi two-dimensional crystal structure well-suited to many transport experiments, while WTe$_2$ and MoTe$_2$ have already been the subject of numerous proposals for device applications. However, with available ARPES techniques it is challenging to demonstrate a Weyl semimetal in \comp. According to the predictions, the Weyl points are above the Fermi level, the system approaches two critical points as a function of doping, there are many irrelevant bulk bands, the Fermi arcs are nearly degenerate with bulk bands and the bulk band gap is small. Here, we study \comp\ for $x = 0.07$ and 0.45 using pump-probe ARPES. The system exhibits a dramatic response to the pump laser and we successfully access states $> 0.2$eV above the Fermi level. For the first time, we observe direct, experimental signatures of Fermi arcs in \comp, which agree well with theoretical calculations of the surface states. However, we caution that the interpretation of these features depends sensitively on free parameters in the surface state calculation. We comment on the prospect of conclusively demonstrating a Weyl semimetal in \comp.
\end{abstract}


\date{\today}
\maketitle

Weyl fermions have been known since the early twentieth century as chiral particles associated with solutions to the Dirac equation at zero mass \cite{Weyl, Peskin}. Although they arise naturally in quantum field theory, they are not known to exist as fundamental particles in nature. However, Weyl fermions are closely related to accidental degeneracies in the band structures of crystals \cite{Herring, Abrikosov, Nielsen, Volovik, Murakami, Multilayer, Pyrochlore, Vish}. By taking advantage of this correspondence, Weyl fermions were observed for the first time, earlier this year, as emergent quasiparticles in the crystal TaAs and in a photonic crystal \cite{TaAsThyUs, TaAsThyThem, TaAsUs, LingLu, TaAsThem}. Such crystals, Weyl semimetals, also host an unusual topological classification that protects topological Fermi arcs on the surface of a bulk sample \cite{Murakami, Multilayer, Pyrochlore}. The bulk Weyl fermions and surface Fermi arcs of a Weyl semimetal are predicted to give rise to a wealth of unusual transport phenomena \cite{Hosur, Potter}. These effects are associated both with the non-trivial topological invariants of a Weyl semimetal and with a variety of phenomena relating to Weyl fermions in quantum field theory. To explore and apply these phenomena, it is important to discover new Weyl semimetals.

Recently, several works have proposed that WTe$_2$, MoTe$_2$ and \comp\ are Weyl semimetals \cite{AndreiNature, TayRong, Binghai, Zhijun}. The \comp\ series may be desirable for transport experiments because it has a highly-layered crystal structure. Indeed, there have already been numerous device proposals for this series, including one proposal for a novel type of transistor based on laser patterning of MoTe$_2$ thin films and another proposal for a topological edge state transistor based on the quantum spin Hall effect in single-layer MoTe$_2$ \cite{LiangFu, LaserPattern}. The intense interest in developing novel devices based on \comp\ suggests that this series may not only provide the first Weyl semimetal outside the TaAs family of compounds, but may also make experimentally accessible novel transport phenomena in Weyl semimetals. In addition, the emergent Weyl fermions in \comp\ are predicted to violate Lorentz invariance, leading to an unexpected and fascinating connection with previously ignored quantum field theories \cite{AndreiNature}.

Despite the intense interest in Weyl semimetals and the unusual properties of \comp, it is exceedingly challenging to experimentally demonstrate a Weyl semimetal in this series. Specifically, WTe$_2$ is close to a critical point for a topological phase transition. Indeed, different theoretical works have predicted that WTe$_2$ is a Weyl semimetal or an insulator by using lattice constants from different X-ray diffraction experiments \cite{AndreiNature, TayRong}. A related concern is the small separation of the Weyl points in momentum space, exceeding the available experimental resolution. Lastly, the Weyl points are above the Fermi level, making them challenging to access using standard angle-resolved photoemission spectroscopy (ARPES). To stabilize the Weyl semimetal phase and increase the momentum-space separation of the Weyl points, it has been proposed to dope WTe$_2$ with Mo \cite{TayRong}. Further, to study the band structure above the Fermi level, it is possible to use pump-probe ARPES. Pump-probe ARPES uses twin lasers to excite electrons above the Fermi level and then perform photoelectron spectroscopy on the excited electrons, giving access to states above the Fermi level. At the same time, the low photon energy afforded by a laser maximizes momentum resolution, ideal for the study of closely-spaced features of the band structure.

Here, we study 7\% and 45\% Mo-doped WTe$_2$ using pump-probe ARPES. We successfully access states $ > 0.2$eV above the Fermi level, well above the energies of the Weyl points \cite{AndreiNature, TayRong}. We find excellent agreement with \textit{ab initio} calculations of surface states. We observe features in the data which may be Fermi arcs. However, we find that the interpretation of the features as Fermi arcs depends sensitively on free parameters in the calculation. We suggest that a careful composition dependence of the \comp\ series can provide more conclusive evidence of Fermi arcs, possibly demonstrating a Weyl semimetal. Our work provides the first step to observing a dramatic violation of Lorentz invariance in the band structure of a crystal.

The crystal structure of \comp\ with an orthorhombic Bravais lattice, space group $Pmn2_1$ ($\#31$) and lattice constants $a = 6.282\textrm{\AA}$, $b = 3.496\textrm{\AA}$, and $c = 14.07\textrm{\AA}$ is shown in Fig. \ref{Fig1}(a) \cite{MoTe2WTe2}. The atomic structure is layered, with single layers of W/Mo sandwiched in-between Te bilayers, which are stacked along the $z$-axis and held together by Van der Waals interactions, see Fig. \ref{Fig1}(a). Seen from Fig. \ref{Fig1}(b), the W/Mo atoms are shifted away from the center of the hexagon formed by the Te atoms. This makes the in-plane lattice constant $a$ longer than the $b$ lattice constant.  Additionally, the distance between adjacent W/Mo atoms is appreciably smaller along $\hat{x}$ than the others, creating high anisotropy. The lack of inversion symmetry is a crucial condition for the Weyl semimetal state in \comp. Shown in Fig. \ref{Fig1}(c) is the scanning electron microscope (SEM) image showing the layered structure of the crystals used in this study. Furthermore, energy-dispersive spectroscopy (EDS) measurements show that the average bulk crystals have composition \half, see Methods. The bulk and (001) surface Brillouin zone (BZ) of \comp\ is shown in Fig. \ref{Fig1}(d). In Fig. \ref{Fig1}(e), we display the bulk band structure of WTe$_2$ along high-symmetry directions. We find that the valence and conduction bands approach each other near $\Gamma$. Therefore, 8 Weyl points are found on the $k_{z}=0$ plane for \twenty, which are formed by the topmost valence and lowest conduction bands, see Fig. \ref{Fig1}(g). In Fig. \ref{Fig1}(f) we note the energy offset between certain Weyl points along the $k_x$ direction. Note, the Weyl points are represented by red and white circles to signify their $+1$ or $-1$ chirality. As shown, two Weyl points are situated $-0.08$eV above $E_{F}$ and two additional Weyl points at $-0.08$eV above $E_{F}$ for \twenty. To convey the difference and uniqueness of the Lorentz-violating Weyl cone in \comp, we compare them with the Lorentz invariant Weyl points recently discovered in the TaAs family of inversion-symmetry breaking single crystals \cite{TaAsThyUs, TaAsThyThem, TaAsUs, LingLu, TaAsThem} in the cartoon schematic shown in Fig. \ref{Fig1}(h). The Lorentz-violating Weyl points, in contrast to the point-like Fermi surface of Lorentz invariant Weyl nodes, have an open Fermi surface. The novel Weyl points in \comp\ grant access to Weyl physics previously ignored in quantum field theory. To resolve the Weyl semimetal state in \half\ we use vacuum ultraviolet angle-resolved photoemission spectroscopy (ARPES) and pump-probe ARPES. In Fig. \ref{Fig1}(i,j) we show the Fermi surface by vacuum ultraviolet APRES on the (001) surface of \half\ along the $\bar{Y}$-$\bar{\Gamma}$-$\bar{Y}$ direction.  The Fermi surface at $E_F$ shows a palmier shaped hole pocket and an almond shaped electron pocket adjacent to the $\bar{\Gamma}$ pocket.

We show that our ARPES spectra agree well with first principles calculations below the Fermi level. This provides an important reference point for understanding the band structure at energies above the Fermi level, where the Weyl points and Fermi arcs are predicted to exist. Figs. \ref{Fig2}(a-c) show the constant energy contours over the $k$-space range defined by the green dotted box in Fig. \ref{Fig1}(j). At the Fermi level, we observe a palmier-shaped pocket and an almond-shaped pocket, see Fig. \ref{Fig2}(a). As one goes from the Fermi level to higher binding energies, the palmier-shaped pocket expands while the almond-shaped pocket shrinks, see Figs. \ref{Fig2}(a-c). This shows that the palmier-shaped pocket is hole-like whereas the almond-shaped pocket is electron-like. We show the energy dispersion along two cuts, namely Cut 1 and Cut 2 defined in Fig. \ref{Fig2}(d). In Cut 1, shown in Figs. \ref{Fig2}(h,i), we observe an electron band and two hole bands at the Fermi level. This is consistent with the fact that Cut 1 goes through the electron pocket and the two convex parts of the hole pocket. In Cut 2, shown in Fig. \ref{Fig2}(j), we observe only the electron band at the Fermi level whereas the hole band disperses to higher binding energies. We compare our Fermi surface with an \textit{ab initio} calculation of the Fermi surface for \twenty. We find that the electron and hole pockets agree well with our experimental data, providing the clearest correspondence yet reported between \textit{ab initio} calculation and ARPES for \comp\ \cite{KaminskiWTe2, FengWTe2, VallaWTe2}. The constant energy contours in Figs. \ref{Fig2}(k-o) show the zoomed-in view of the region of approach between the electron and hole pockets. We see that, as one goes from higher binding energies to the Fermi level, the two pockets gradually approach each other. At the Fermi level, shown in Fig. \ref{Fig2}(k), the separation between the two pockets is small but finite. We show a series of the band dispersions in Figs. \ref{Fig2}(p-s), where, again, we observe both the electron and the hole bands. It is important to note that the ARPES data are dominated by the contribution from surface states, or more precisely, states which are mostly localized on the surface and whose spectral weight mainly comes from the surface region. One piece of evidence is that our constant energy contour data in Figs. \ref{Fig2}(a-c) show excellent agreement with the theoretical calculations in Figs. \ref{Fig2}(e-g), which only considers the spectral weight from the top unit cell of a semi-infinite system. In addition, the observed bands are sharp. If the spectral weight were due to the bulk bands, we expect the contours to be filled in with intensity due to the finite $k_z$ resolution. Lastly, we note that the sharpest states in our ARPES spectra, the almond pocket near the region where it intersects the palmier, correspond to a state heavily localized near the surface in the calculate, indicated by a heavy red region in Figs. \ref{Fig2}(e,g). We provide an illustration of the measured Fermi surface in Fig. \ref{Fig2}(v). According to the theoretical calculations, as one goes to energies above the Fermi level, the two pockets are expected to further approach each other. Eventually the convex region of the palmier pocket intersects the almond pocket, giving rise to Weyl nodes.




Next, we show that the almond pocket in \half\ evolves into two nested electron pockets above the Fermi level. To access the unoccupied states, we use pump-probe ARPES with a $1.49$eV pump laser to excite electrons into low-lying states above the Fermi level and a $5.99$eV probe laser to perform photoemission \cite{IshidaMethods}. We can, further, directly and precisely control the time delay between the two lasers. Here, we suppose that the pump laser arrives always at time $t_0$ and we study ARPES spectra at two probe times $t$. For $t < t_0$, the probe arrives before the pump, so we only observe occupied states, as in normal ARPES. For $t > t_0$, the probe arrives after the pump and we observe unoccupied states. We use pump probe ARPES with $t > t_0$ to understand the band structure in \comp\ above the Fermi level. Corresponding spectra with $t < t_0$ serve as a reference to compare with normal ARPES. We first present a cut of \half\ along the $\bar{\Gamma}$-$\bar{Y}$ direction at $k_x = 0$, see Figs. \ref{Fig3}(a,b). The sample responds beautifully to the pump laser and we observe a dramatic evolution of the bands up to energies $> 0.2$eV above the Fermi level. Note that all available calculations of \comp\ place the Weyl points $< 0.1$eV above the Fermi level. In addition, the Weyl points are all predicted to be within $0.25 \textrm{\AA}^{-1}$ of the $\bar{\Gamma}$ point \cite{AndreiNature, TayRong, Binghai, Zhijun}. We see that our pump probe measurement easily accesses the relevant region of reciprocal space. We see in Fig. \ref{Fig3}(b) that the palmier pocket closes up at $\sim -0.02$eV of the Fermi level. However, the almond pocket extends well above the Fermi level. In particular, we notice two branches reaching toward $\bar{\Gamma}$ and appearing to merge at $\sim -0.15$eV. These bands correspond to the inner and outer contours of the almond, discussed above. We also see a broader, diffuse branch of the almond pocket reach toward the $\bar{Y}$ point. We find that above $E_F$, the palmier closes up rapidly while the almond expands outward into a large pair of electron pockets. To better understand this evolution, we show a Fermi surface mapping of the almond pocket above $E_F$, see Fig. \ref{Fig3} (d). We see the almond pocket grows larger above $E_F$, again, as an electron pocket. We also find that the almond pocket evolves into two nested contours. This structure is most easily visible at $E_\textrm{B} = -0.1$eV in Fig. \ref{Fig3} (d). We reproduce this constant energy cut in Fig. \ref{Fig3} (h), where we also mark the two contours by blue dotted lines. From the $\bar{\Gamma}$-$\bar{Y}$ cut, we see that this pair of nested contours corresponds to the two contours which make up the almond pocket at $E_F$, as shown in Fig. \ref{Fig2} (a). Next, we study the band structure on a cut parallel to $\bar{\Gamma}$-$\bar{X}$ at $k_y \sim 0.36 \textrm{\AA}^{-1}$, see Figs. \ref{Fig3}(e,f). Again we see a pair of electron bands which we associate with the almond pocket. The lower branch, marked (1) in Fig. \ref{Fig3}(g), corresponds to the sharp, prominent branch of the almond as it appears below $E_F$. On the other hand, the upper branch, (2) of the almond pocket has a minimum at $\sim -0.02$eV on this cut and is entirely above the Fermi level. In addition, we note that this upper branch appears to have a discontinuity at $k_x \sim 0.02 \textrm{\AA}^{-1}$ where it seems to break into two disjoint bands, (2) and (3). It is this feature which is our candidate Fermi arc and which we analyze at greater length below. For reference, in Fig. \ref{Fig3} (i) we label the cuts shown in Figs. \ref{Fig3} (a,b,e,f).

Next, we show that the disjoint band we observe agrees well with a topological Fermi arc found by first principles calculation. We consider a cut of \half, shown in Fig. \ref{Fig4}(a) at $k_y = 0.325 \textrm{\AA}^{-1}$, with \textit{ab initio} calculations taken at $k_y = k_\textrm{W1} = 0.215 \textrm{\AA}^{-1}$, Fig. \ref{Fig4}(b-e). We note that the $k_y$ values differ, but both correspond to the edge of the electron pocket at the Fermi level, the approximate location of the Weyl points. In addition, as we show below, the band structure features we discuss here appear robust as a function of doping. As noted above, the discrepancy may be due to an error in determining $\bar{\Gamma}$ in experiment, an error in the approximation used for calculating doped band structures, or, especially, a systematic error due to comparing a calculation of $20\%$ doping with a measurement of $45\%$ doping. We further show calculations for different on-site surface potentials, $0$eV and $-0.11$eV, in Fig. \ref{Fig4}(b,c), respectively. We find that the two surface potentials give rise to a different connectivity pattern of the Fermi arcs \cite{TayRong}. In particular, for $0$eV, we find that two short arcs connect neighboring Weyl points, as shown in Fig. \ref{Fig4}(d,j), while for $-0.11$eV, one arc connects the two $W_1$ and a long arc emanates from the $W_2$, as shown in Fig. \ref{Fig4}(e,k). The surface potential is difficult to predict, but can be fixed by comparison with ARPES data. Comparing our ARPES spectrum of \half\ with the numerics, we see that our ARPES spectrum suggests that the two $W_1$ are connected by a Fermi arc, favoring the scenario in Fig. \ref{Fig4}(k). This suggests that we have observed Fermi arcs. We also find that the energy separation of the arc features is $\sim 0.03$eV, consistent with the energy separation of the Weyl points. Similarly, we can estimate the momentum separation between the two arc features to be $\sim 0.03 \textrm{\AA}^{-1}$, consistent with \textit{ab initio} calculation \cite{TayRong}. However, the interpretation of the features in the data is complicated by the trivial surface states. First, while the $-0.11$eV calculation shows two trivial electron pockets in addition to the Fermi arc, in our ARPES data we only observe in total two electron pockets. Furthermore, the $0$eV calculation shows trivial surface states which agree well with the flat trivial surface state at $\sim -0.1$eV that merges into the bulk, giving rise to a trivial surface state arc, see Fig. \ref{Fig4}(b). This cut further shows a trivial surface state electron pocket with branches which merge into the bulk near the flat surface state. These features also match well with the ARPES spectrum, but suggest that we see only trivial surface states. Moreover, in this case, the Fermi arcs would be nearly degenerate with adjacent trivial surface states, making them very difficult to observe given the typical linewidth of our spectra. We further study a cut of \ten\ at $k_y = 0.29 \textrm{\AA}^{-1}$. For these spectra, we find that $p$ polarization selects the Fermi arc feature, see Fig. \ref{Fig4}(f), while the $s$ polarization gives two electron pockets, see Fig. \ref{Fig4}(g). This suggests a total of three electron bands, which matches the $-0.11$eV case. However, again we can also interpret the bands as trivial pockets which appear as arcs because they merge into the bulk. Lastly, we also note that the center arc, labelled (2) in Fig. \ref{Fig4}(i) is above the outer arcs (1) in \ten, while arc (3) is below arc (2) in Fig. \ref{Fig2}(g). This evolution of the structure as a function of composition provides one route to clarify whether or not the disjoint feature we observe in \half\ and \ten\ is a topological Fermi arc. By comparing our experimental data with \textit{ab initio} calculation, we find that the disjoint band we observe in our ARPES spectra agrees with topological Fermi arcs in numerics, providing the first direct experimental signature of a topological Fermi arc in \half\ and \ten. 

We discuss the extent to which the signatures of Fermi arcs observed here are robust. In support of an observation of Fermi arcs: (1) we observe surface states that are disjoint in their dispersion; (2) the energy spacing between the terminations of disjoint surface states also agrees with the energy difference between the W1 and W2 Weyl nodes in calculations; (3) the momentum separation of the disjoint surface state agrees within error with the separation of the Weyl nodes in calculation; (4) the overall experimentally-measured dispersions match with the calculations. However, it is important to note that, in general, a qualitative agreement between the experimental data and the theoretical calculations is not a conclusive demonstration of the Weyl semimetal state. In particular, we see that the calculated dispersion changes dramatically as a function of the on-site potential, see again Figs. \ref{Fig4}{\bf b,c}. As importantly, the disjoint nature of the surface states is a necessary, but not sufficient condition for showing Fermi arcs. We see that in the calculation there are trivial surface states that are not Fermi arcs but also show a disjoint dispersion. This is not surprising since a trivial state may merge into a bulk band, forming a trivial arc. This effect is particularly relevant for \comp\ because there are many irrelevant bulk bands at the energies of the Weyl points.

We compare the experimental results here with those in the TaAs class to further elaborate on the prospect of conclusively demonstrating Fermi arcs in \comp. We have experimentally demonstrated the existence of Fermi arcs in TaAs and TaP by counting the net number of chiral surface states along a $k$-space loop that encloses a projected Weyl node. There, the demonstration did not rely directly on theoretical calculations and was robust. The following properties were important for achieving this robust proof in TaAs and TaP \cite{TaAsUs, TaPUs, NbPUs}: (1) The $k$-space separation between the W2 Weyl nodes in TaAs and TaP is sufficiently large, $\Delta{k} \sim 0.08 \textrm{\AA}^{-1}$. (2) Compared to \comp, TaAs and TaP are nearly ideal semimetals, with only small irrelevant bulk bands at the energies of the Weyl nodes. Also, the bulk band structure including both the Weyl cones and the trivial bands were experimentally determined by soft X-ray ARPES. Combining these two conditions, it was possible to find a $k$-space loop that encloses a projected Weyl node, and, at the same time, along which the bulk band structure has a full energy gap. Now, we evaluate the situation in \comp\ also based on these two conditions. (1) The $k$-space separation between the Weyl nodes is predicted to be $\sim 0.04 \textrm{\AA}^{-1}$ for \half\ \cite{TayRong}. This separation, although smaller than the separation of Weyl points in TaAs, is still within the momentum resolution of ARPES. (2) \comp\ is highly metallic, with many irrelevant bulk bands. Moreover, the energy gap between the two bands that cross to form the Weyl nodes is small. For example, the band gap is $<15$ meV at all $k$ points along the cut connecting adjacent $W_1$ and $W_2$ \cite{TayRong}. For these reasons, it is difficult to demonstrate the Fermi arcs in \comp\ by a direct measurement of a non-zero Chern number, as can be carried out in TaAs and TaP. To conclusively demonstrate a Weyl semimetal in \comp\ it is necessary, instead, to observe an isolated disjoint contour. We propose to study the evolution of the candidate Fermi arc with composition, $x$. A switch in the connectivity at a special composition or a composition where the Fermi arc feature is well-isolated from irrelevant bands may provide a conclusive proof that \comp\ is a Weyl semimetal.

\ \\ 
{\bf Methods}
\\ 

Synchrotron-based ARPES measurements were performed at the CASSIOPEE beamline at Soleil in Saint-Aubin, France and the I05 beamline at the Diamond Light Source (DLS) in Didcot, United Kindom. ARPES measurements were also carried out using a home-built laser-based ARPES setup at the Ames Laboratory in Ames, Iowa, United States. All measurements were conducted under ultra-high vacuum and at temperatures $\leq 10$K. The angular and energy resolution of the synchrotron-based ARPES measurements was better than $0.2^{\circ}$ and $20$ meV, respectively. Photon energies from 15 eV to 100 eV were used. The angular and energy resolution of the home-built laser-based ARPES measurements was better than $0.1^{\circ}$ and $5$ meV, respectively. Photon energies from 5.77 eV to 6.67 eV were used.

The pump-probe ARPES apparatus consisted of a hemispherical analyzer and a mode-locked Ti:Sapphire laser system that delivered $1.48$ eV pump and $5.92$ eV probe pulses at a $250$kHz repetition \cite{IshidaMethods}. The time and energy resolution was $300$ fs and $27$ meV, respectively. The spot diameter of the pump and probe beams at the sample position was $250$ and $85 \mu$m, respectively. Samples were cleaved in the spectrometer at $<5 \times 10^{-11}$ Torr, and measurements were conducted at $\sim8$ K.

High quality ribbon-like single crystals of Mo$_x$W$_{1-x}$Te$_2$ were grown by chemical vapor transport (CVT) with iodine (I) as the agent. Before growing the crystals, the quartz tubes were thoroughly cleaned using thermal and ultrasonic cleaning treatments to avoid contamination. Stoichiometric amounts of W (99.9\% powder, Sigma-Aldrich), Mo (99.95\% powder, Sigma-Aldrich) and Te (99.95\%, Sigma-Aldrich) were mixed with iodine, were sealed in a $20$ cm long quartz tube under vacuum $\sim 10^{?6}$ Torr, and then placed in a three-zone furnace. The reaction zone dwelled at $850$ $^{\circ}$C for 40 hours with the growth zone at $900$ $^{\circ}$C and then heated to $1070$ $^{\circ}$C for seven days with the growth zone at $950$ $^{\circ}$C. Lastly, the furnace was allowed to cool naturally down to room temperatures. The Mo$_x$W$_{1-x}$Te$_2$ single crystals were collected from the growth zone. Excess iodine adhering to the single crystals was removed by using acetone or ethanol. Nominal compositions of the Mo$_x$W$_{1-x}$Te$_2$ crystals were $x = 0.0, 0.1, 0.15, 0.2, 0.3$. An energy dispersive spectroscopy (EDS) measurement was carried out to precisely determine the composition of the samples. Samples were first surveyed by an FEI Quanta 200FEG environmental scanning electron microscope (SEM). The chemical compositions of the samples were characterized by an Oxford X-Max energy dispersive spectrometer that was attached to the SEM. All the samples were loaded at once in the SEM chamber to ensure a uniform characterization condition. Both the SEM imaging and EDS characterization were carried out at an electron acceleration voltage of $10$ kV with a beam current of $0.5$ nA. For each sample, three different spatial positions were randomly picked to check the uniformity. The measured compositions were: MoTe$_2$ for $x = 0$, Mo$_{0.42}$W$_{0.58}$Te$_2$ for $x= 0.1$,  Mo$_{0.46}$W$_{0.54}$Te$_2$ for $x = 0.15$, Mo$_{0.44}$W$_{0.56}$Te$_2$ for $x = 0.2$ and Mo$_{0.43}$W$_{0.57}$Te$_2$ for $x = 0.3$. For $x \neq 0$, we considered the samples to all have the approximate composition Mo$_{0.45}$W$_{0.55}$Te$_2$. Single crystals of Mo$_{0.07}$W$_{0.93}$Te$_2$ were grown by the self-flux technique. Raw materials in the molar ratio of $7:93:200=\textrm{Mo}:\textrm{W}:\textrm{Te}$ were sealed in a quartz tube under vacuum. The mixture was heated to $750$ $^{\circ}$C, held for 1 day, and then quenched in cold water. The obtained Mo$_x$W$_{1-x}$Te$_2$ powder ($150$ mg) and $20$ g Te were placed in a new quartz tube and sealed under vacuum. The tube was heated to $900$ $^{\circ}$C, held for $1$ day and then cooled down to $500$ $^{\circ}$C over $10$ days. At $500$ $^{\circ}$C, the Te flux was separated by centrifugation. The crystal structure was determined by a Bruker SMART diffractometer as previously reported in Ref. \cite{SongMethods}. Electrical transport measurements were carried out in a homemade system with a minimum temperature of $2$ K and a maximum field of $9$ Tesla. Ohmic contacts were made using gold wires and silver paste. The observations of Shubnikov de haas quantum oscillations and large magnetoresistance indicated the high quality of the samples.

We computed the electronic structures by using the projector augmented wave method \cite{PAW1,PAW2} as implemented in the VASP \cite{TransitionMetals, PlaneWaves1, PlaneWaves2, GGA} package within the generalized gradient approximation (GGA) schemes \cite{GGA}. For WTe$_2$, the experimental lattice constants used were from \cite{MoTe2WTe2}. A $8 \times 16 \times 4$ Monkhorst Pack $k$-point mesh was used in the computations. The lattice constants and atomic positions of MoTe$_2$ were fully optimized in a self-consistent calculation for an orthorhombic crystal structure until the force became less than $0.001$ eV$/\textrm{\AA}$. Spin-orbit coupling was included in our calculations. To calculate the bulk and surface electronic structures, we constructed a first-principles tight-binding model Hamiltonian for both WTe$_2$ and MoTe$_2$, where the tight-binding model matrix elements were calculated by projecting onto the Wannier orbitals \cite{MLWF1,MLWF2,Wannier90}, which used the VASP2WANNIER90 interface \cite{MLWF3}.  We used the $s$- and $d$-orbitals for W(Mo) and the $p$-orbitals for Te to construct Wannier functions, without performing the procedure for maximizing localization. The electronic structure of Mo$_x$W$_{1-x}$Te$_2$ samples was calculated by a linear interpolation of the tight-binding model matrix elements of WTe$_2$ and MoTe$_2$. The surface states were calculated by the surface Green's function technique \cite{Green}, which computed the spectral weight near the surface of a semi-infinite system.

\ \\
{\bf Acknowledgements}
\\

I.B. and D.S. thank Moritz Hoesch and Timur Kim for support during synchrotron ARPES measurements at Beamline I05 of Diamond Lightsource in Didcot, UK. I.B. acknowledges the support of the US National Science Foundation GRFP. X.C.P., Y.S., B.G.W., G.H.W and F.Q.S. thank the National Key Projects for Basic Research of China (Grant Nos. 2013CB922100, 2011CB922103), the National Natural Science Foundation of China (Grant Nos. 91421109, 11522432, and 21571097) and the NSF of Jiangsu province (No. BK20130054).

\clearpage
\begin{figure}
\centering
\includegraphics[scale=0.68, trim={0 10 0 10}, clip]{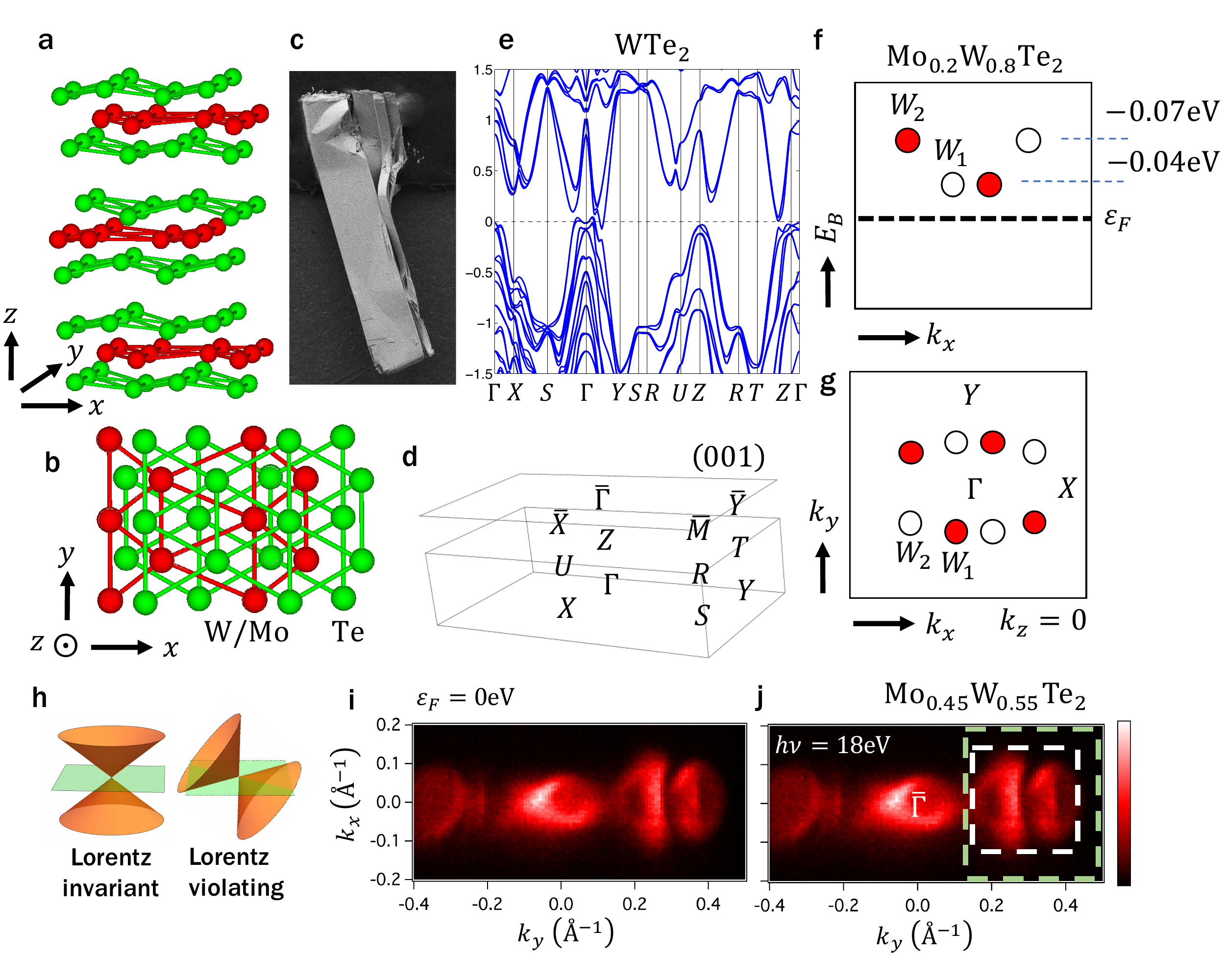}
\end{figure}

\clearpage
\begin{figure*}
\caption{\label{Fig1}\textbf{Overview of Mo$_x$W$_{1-x}$Te$_2$.} {\bf a,} Side view of the Mo$_x$W$_{1-x}$Te$_2$ crystal structure. Mo$_x$W$_{1-x}$Te$_2$ crystallizes in a orthorhombic structure with space group $Pmn2_1$ ($\#31$). The W/Mo atom layers are shown in red and the Te atom layers in green. {\bf b,} Top view of (a), showing only one monolayer, consisting of a layer of of W/Mo atoms sandwiched between two layers of Te atoms. The structure lacks inversion symmetry. {\bf c,} A beautiful scanning electron microscope (SEM) image of a $\sim{0.4 \times 2}$ mm rectangular shaped \half\ bulk crystal clearly demonstrates a layered structure. {\bf d,} The bulk and (001) surface Brillouin zone (BZ) of \comp\ with high symmetry points labeled. {\bf e,} The electronic bulk band structure of WTe$_2$. There is a band crossing near $\Gamma$ which gives rise to electron and hole pocket along the $\Gamma$-$Y$ direction. {\bf f,} {\bf g,} According to calculation, \twenty\ has two sets of Weyl points, $W_1$ and $W_2$, at different energies. In total, there are  eight Weyl points, all on the $k_{z}=0$ plane. The red and white circles represent Weyl points with $+1$ or $-1$ chirality. {\bf h,} Lorentz-invariant and Lorentz violating Weyl cones. Unlike Lorentz invariant Weyl cones, Lorentz-violating Weyl cones are tilted over, and have a constant energy contour with an electron and hole pocket touching at a point. {\bf i,} {\bf j,} Fermi surface by vacuum ultraviolet APRES on the (001) surface of \half\ at $\varepsilon_{F}=0$eV with an incident photon energy of 18 eV. Adjacent to the $\bar{\Gamma}$ pocket we observe a palmier-shaped hole pocket and an almond-shaped electron pocket along the $\bar{Y}$-$\bar{\Gamma}$-$\bar{Y}$ direction. The region enclosed by green dashed box corresponds to Fig. \ref{Fig2} (a)-(d), and that surrounded by the white dashed box corresponds to Fig. \ref{Fig2} (k)-(o). We expect to see Weyl points and Fermi arcs above the Fermi level in the region where the electron and hole pockets intersect.}
\end{figure*}

\clearpage
\begin{figure}
\centering
\includegraphics[scale=0.7, trim={0 10 10 10}, clip]{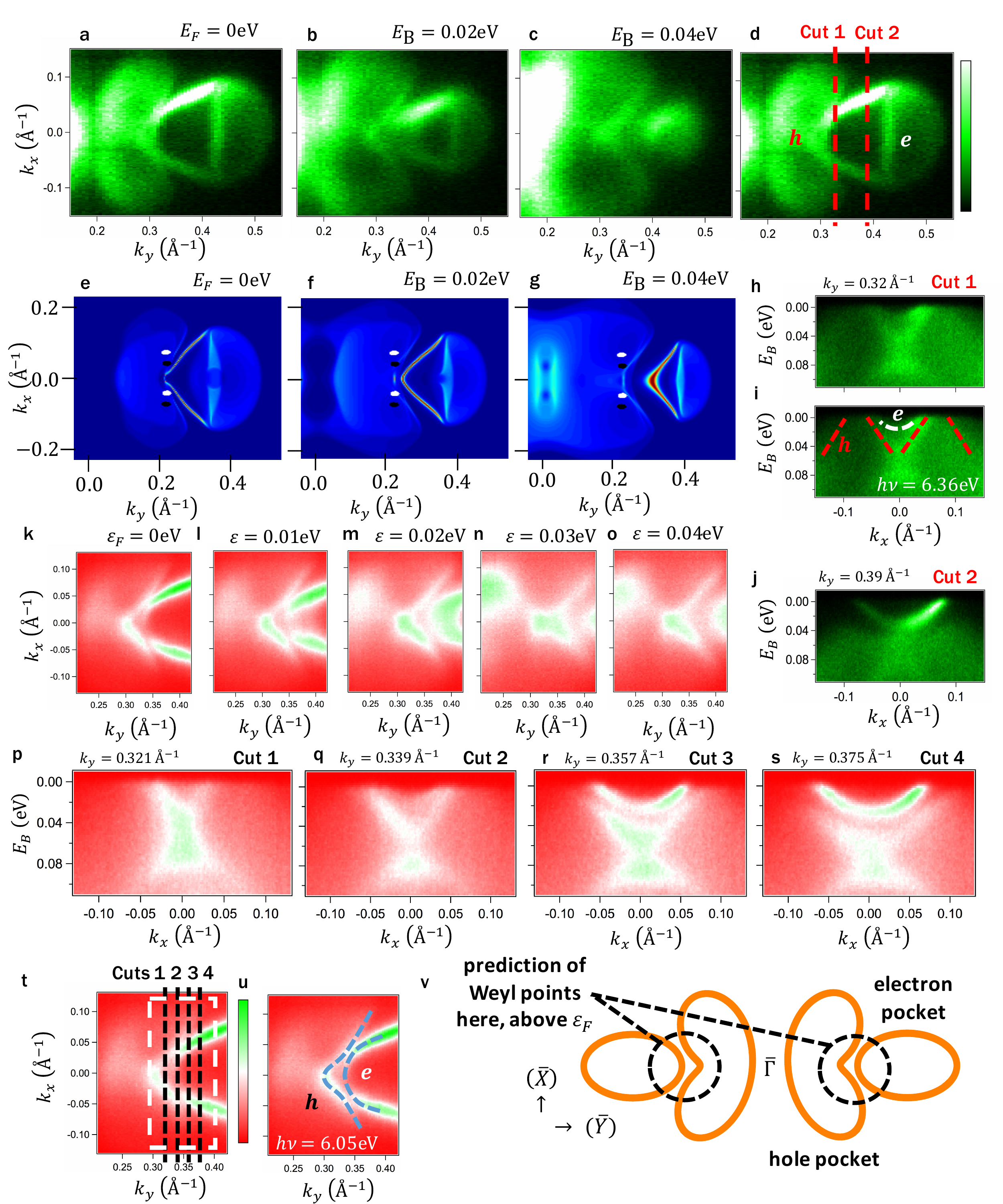}
\end{figure}

\clearpage
\begin{figure*}
\caption{\label{Fig2}\textbf{Mo$_x$W$_{1-x}$Te$_2$ below the Fermi level.} {\bf a-c,} ARPES measured constant energy contour maps at different binding energies using the photon energy $h\nu=6.36$ eV. The $k$-space range of the maps is defined by the green dotted box in Fig. 1j. We identify a palmier-shaped hole-like pocket and an almond-shaped electron pocket. {\bf d,} The same data as panel (a). The electron and hole pockets are labeled by $e$ and $h$, respectively. {\bf e-g,} First-principles calculated constant energy contours at different binding energies. {\bf h,} ARPES measured $E_{\textrm{B}}-k_y$  dispersion map along Cut 1. {\bf i,} The same data as panel (h). The dotted lines are guides to the eye that roughly trace the dispersion of the bands. {\bf j,} ARPES measured energy dispersion map along Cut 2. Cut 1 and Cut 2 are defined by the red lines in panel (d). {\bf k-o,} ARPES measured constant energy contour maps at different binding energies using the photon energy $h\nu=6.05$ eV. The $k$-space range of the maps is defined by the white dotted box in Fig. 1j. {\bf p-s,} ARPES measured $E_{\textrm{B}}-k_y$ dispersion maps along the cuts that are defined by the black dotted lines in panel (t). {\bf t,u,} The same data as panel (p). The blue dotted lines in panel (u) are guides to the eye that roughly trace the electron and hole pockets at the Fermi level. {\bf v,} Schematic illustration of the measured Fermi surface. As one goes to energies above the Fermi level, the electron and hole pockets are expected to further approach each other. Eventually the convex parts of the palmier-shaped hole pocket will touch the electron pocket, forming Weyl nodes. The dotted circles show the $k$-space region, in which the Weyl nodes are expected.}
\end{figure*}

\begin{figure}
\centering
\includegraphics[scale=0.4, trim={10 10 10 10}, clip]{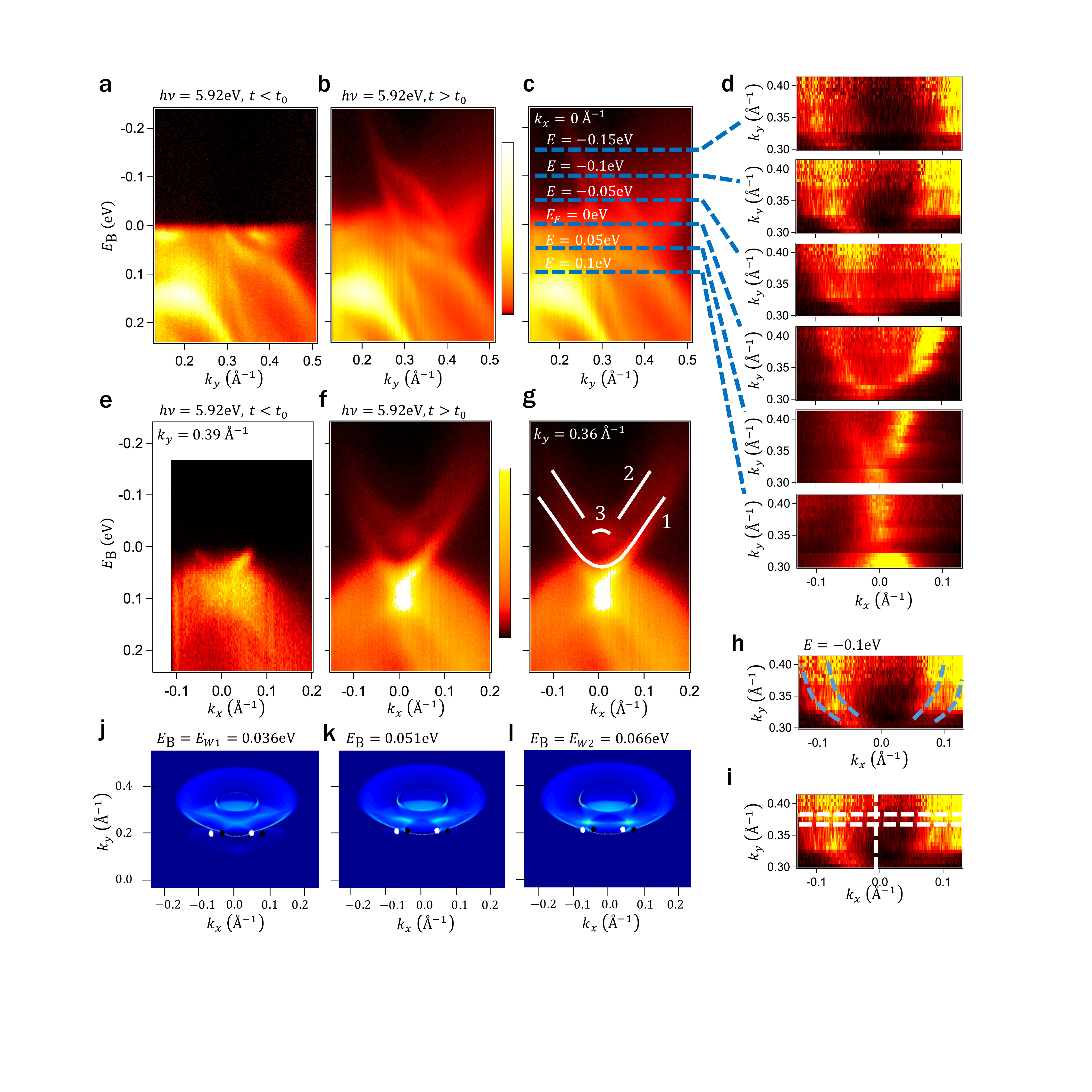}
\end{figure}

\clearpage
\begin{figure*}
\caption{\label{Fig3}\textbf{\comp\ above the Fermi level.} {\bf a, b,} We first present an energy dispersion cut of \half\ along the $\bar{\Gamma}$-$\bar{Y}$ direction at $k_{x} = 0$. For $t < t_{0}$, the probe arrives before the pump, so we see only states below the Fermi level, while for $t > t_{0}$ the probe arrives after the pump and we observe states above the Fermi level. We see $> 0.2$eV above $E_{F}$, well above the energies of the Weyl points. {\bf c,} Same as (b) but with blue dotted lines at various energies. {\bf d,} The evolution of the Fermi surface for the almond pocket at the energy levels marked in (c). Here we observe that the almond pocket evolves into two nested contours, seen most clearly at $E_{B}=-0.1$eV. {\bf e, f,} Energy dispersion of \half\ for $t > t_0$ and $t < t_0$ on a cut parallel to $\bar{\Gamma}$-$\bar{X}$ at $k_{y}\sim 0.36 \textrm{\AA}$. For $t < t_0$ we see a single electron band, associated with the almond pocket. For $t < t_0$, we find an additional electron band with minimum at $\sim -0.02$eV. We observe a discontinuity in the band at $\sim \pm 0.02 \textrm{\AA}^{-1}$. {\bf g,} Same as (f), with guides to the eye to highlight the discontinuity in the electron pocket above the Fermi level. The candidate Fermi arcs are features (2) and (3). {\bf h,} At $E_{B}=-0.1$eV, the two nested contours are marked by blue dotted lines. {\bf i,} Cuts shown in (a), (b), (e), and (f) are labeled for reference. {\bf j-l,} Calculation of the Fermi surface for \twenty\ above $E_F$. We see an electron pocket, consistent with the measured Fermi surface above $E_F$, (d).}
\end{figure*}

\begin{figure}
\centering
\includegraphics[scale=0.45, trim={0 0 0 0}, clip]{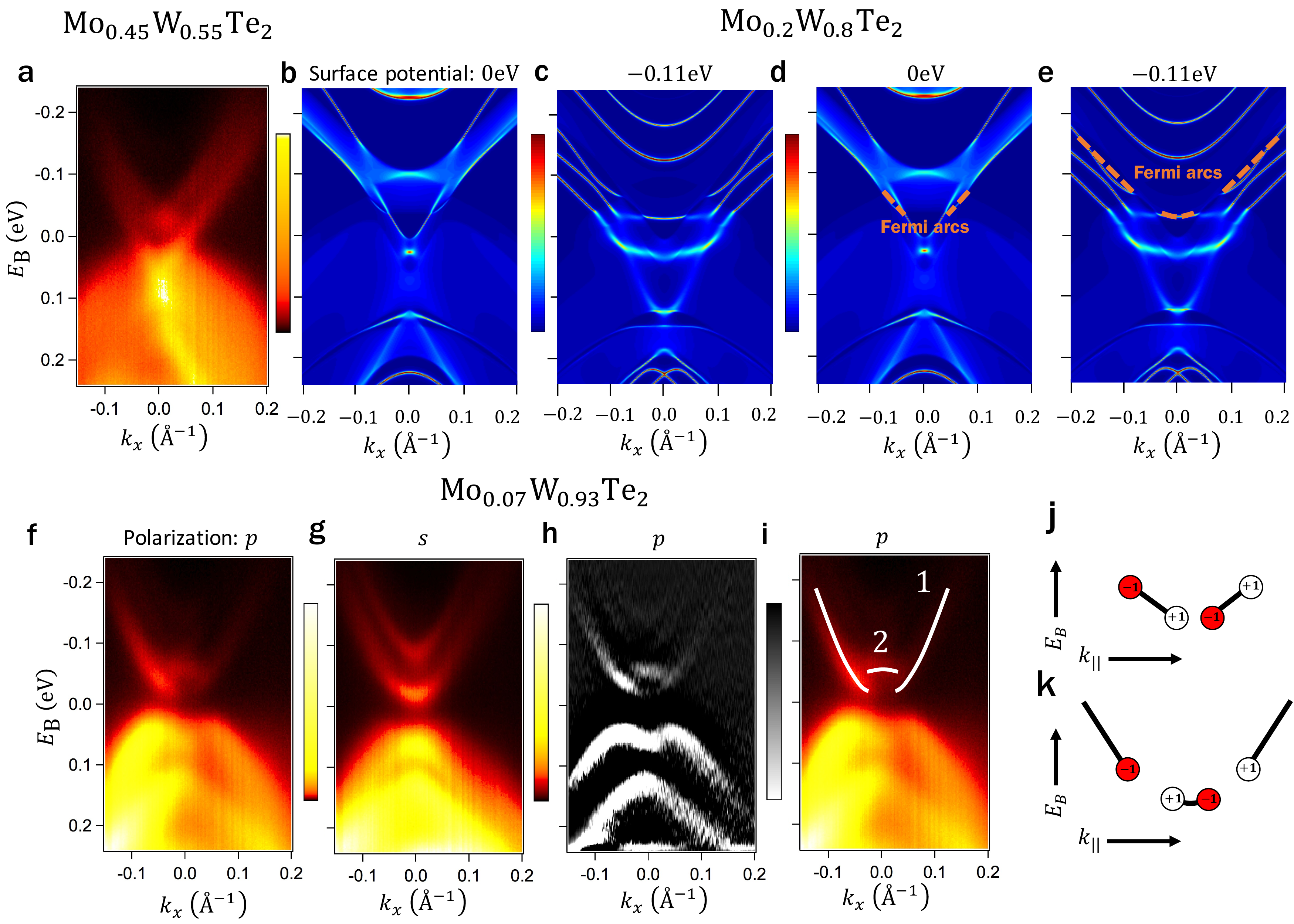}
\end{figure}

\begin{figure*}
\caption{\label{Fig4}\textbf{Signatures of Fermi arcs in \comp.} {\bf a,} Pump-probe ARPES spectrum at $k_y = 0.325 \textrm{\AA}^{-1}$, showing disjoint features, as marked in Fig. \ref{Fig3} (g). {\bf b,} \textit{Ab initio} calculation at $k_y = k_\textrm{W1} = 0.215 \textrm{\AA}^{-1}$ with $0$eV surface potential, showing trivial surface states which merge into the bulk bands, forming trivial arcs. The trivial disjoint features match the data well. {\bf c,} Same calculation but with $-0.11$eV surface potential, showing topological Fermi arcs which match the data well. {\bf d, e,} Same as (b, c) but with the arc marked by dotted orange lines. Note the many trivial surface states, many of which merge into the bulk, forming trivial arcs. The proliferation of trivial surface states complicates interpretation of the ARPES spectra. {\bf f,} Pump-probe ARPES spectrum of \ten\ at $k_y = 0.29 \textrm{\AA}^{-1}$ and polarization $p$, emphasizing the disjoint feature and {\bf g,} polarization $s$, showing two trivial electron surface states. Like \half, we the disjoint features consist of an inner arc and two outer arcs, as shown in {\bf h,} a second derivative plot and {\bf i,} marked on the data. Note that the inner arc is above the ends of the outer arcs in \ten, while the inner arc is below the outer arcs in \half. This composition dependence suggests one route to clarifying whether or not we observe topological Fermi arcs. {\bf j,} The connectivity of Fermi arcs we find with surface potential $0$eV, with a $W_1$ connected to a $W_2$ and {\bf k,} the connectivity for $-0.11$eV, with the two $W_1$ connected and two $W_2$ connected. This second scenario is suggested by our data.}
\end{figure*}

\end{document}